\documentstyle[12pt]{article}
\begin{document}
\begin{center}{\large{\bf Mapping among manifolds 1}}
\end{center}
\vspace*{1.5cm}
\begin{center}
A. C. V. V. de Siqueira
$^{*}$ \\
Departamento de Educa\c{c}\~ao\\
Universidade Federal Rural de Pernambuco \\
52.171-900, Recife, PE, Brazil\\
\end{center}
\vspace*{1.5cm}
\begin{center}{\bf Abstract}

In this paper we have build the modified Hamiltonian formalism for
geometric objects like the Jacobi fields and metric tensors. In
this approach Jacobi fields and metric tensors are mapped among
manifolds. As an application, we have mapped a general
n-dimensional Riemannian manifold to a n-dimensional maximally
symmetric spacetime.
\end{center}

 \vspace{3cm}

${}^*$ E-mail: acvvs@ded.ufrpe.br
\newline

\newpage

\section{Introduction}
$         $

The Jacobi fields are very important to the Riemannian Geometry
[1] and in the singularity theorems [2],[3]. These fields were
used to study a free falling particle motion in a Schwarzschild
spacetime [4], and a charged particle motion in Kaluza-Klein
manifolds [5]. In this paper we present a map building method
among manifolds. As an application, we have chosen, for their
importance, a map between a n-dimensional general Riemannian
manifold and a n-dimensional maximally symmetric spacetime.

 This paper is organized as follows. In Sec. $2$ we present some facts
about the Jacobi fields. In Sec. $3$  we build the Jacobi equation
in a vielbein basis. In Sec. $4$ we modify the Hamiltonian
formalism to a new version, that we call \emph{the modified
Hamiltonian formalism}. In Sec. $5$ we build a map among manifolds
by the use of the modified Hamiltonian formalism for the Jacobi
fields on a geodesic curve.. In Sec. $6$ we apply the modified
Hamiltonian formalism to the Jacobi fields on a non-geodesic
curve. In Sec. $7$ we apply the modified formalism to  metric
tensors. In Sec. $8$ we summarize the main results of this work.

\renewcommand{\theequation}{\thesection.\arabic{equation}}
\section{\bf Jacobi Fields}
\setcounter{equation}{0}
$         $

In this section we briefly review the Jacobi fields and their
respective differential equation for a Riemann manifold. Let us
consider a differentiable manifold, $\cal{M}$, and two structures
defined on $\cal{M}$, namely affine connection, $\nabla$ , and
Riemann tensor, $K$ , related by the equation [2],[3]

\begin{equation}
 K(X,Y)Z=\nabla _{X}\nabla _{Y}Z-\nabla _Y\nabla
_{X}Z-\nabla _{[X,Y]}Z,
\end{equation}
where $X,Y$ and $Z$ are vector fields in the tangent space. The
torsion tensor can be defined by
\begin{equation}
 T(X,Y)=\nabla _{X}Y-\nabla _{Y}X-[X,Y],
\end{equation}
where $\nabla _{X}Y$ is the covariant derivative of the field $Y$
along the $X$ direction. We  define the Lie derivative by

\begin{equation}
 {\cal{L}}_{X}Y=[X,Y].
\end{equation}
For a torsion-free connection we can write

\begin{equation}
 {\cal{L}}_{X}Y=\nabla _{X}Y-\nabla _{Y}X.
\end{equation}
Let us consider the tangent field V, the Jacobi field Z and the
condition
\begin{equation}
 {\cal{L}}_{V}Z=0.
\end{equation}
From (2.5), we can easily see that the equations which govern the
Jacobi field can be written in the form

\begin{equation}
 \nabla _V\nabla _VZ+K(Z,V)V-\nabla _Z\nabla _VV=0.
\end{equation}
 In this section and in sections $3$ and $5$ we assume the simplest condition
\begin{equation}
 \nabla _{V}V=0.
\end{equation}
In this case the Jacobi equations are reduced to a geodesic
deviation, assuming a simpler form, and the Fermi derivative
$\frac{D_FZ}{\partial s}$ coincides with the usual covariant
derivative

$$
\frac{D_FZ}{\partial s}=\frac{DZ}{\partial s}=\nabla _VZ,
$$
and

\begin{equation}
 \nabla _V\nabla _VZ+K(Z,V)V=0.
\end{equation}

\renewcommand{\theequation}{\thesection.\arabic{equation}}
\section{\bf Jacobi Equation in a Vielbein Basis}

\setcounter{equation}{0}
$         $

We intend to make use of a vielbein basis related to a local
metric G, as usual. The metric tensor $G_{\Lambda\Pi}$ has
signature $(-,+,\ldots,+),$ with curved indices $\Lambda,\Pi$
$\in$ $(0,1,2,3,4,5,6,\ldots,n)$ and it is associated to a local
coordinate basis. Let us consider the connection between the
vielbein and the local metric tensor
\begin{equation}
G_{\Lambda\Pi}=E_{\Lambda}^{(\mathbf{A})}E_{\Pi}^{(\mathbf{B})}\eta_{(\mathbf{A})(\mathbf{B})},
\end{equation}
where $ \eta_{(\mathbf{A})(\mathbf{B})}$ and $
E_{\Lambda}^{(\mathbf{A})}$ are Lorentzian metric and vielbein
components respectively. The flat indices
$(\mathbf{A}),(\mathbf{B}),\ldots,(\mathbf{M}),(\mathbf{N})$ $\in$
$(0,1,2,3,4,5,6,\ldots,n).$
 In the vielbein basis we use the following Riemannian curvature
\begin{eqnarray}
&& K_{(\mathbf{A})(\mathbf{B})(\mathbf{C})(\mathbf{D})}=-\gamma_{(\mathbf{A})(\mathbf{B})(\mathbf{C}),(\mathbf{D})}+\gamma_{(\mathbf{A})(\mathbf{B})(\mathbf{D}),(\mathbf{C})}\\
\nonumber
&& +\eta^{(\mathbf{M})(\mathbf{N})}[\gamma_{(\mathbf{B})(\mathbf{A})(\mathbf{M})}(\gamma_{(\mathbf{C})(\mathbf{N})(\mathbf{D})}-\gamma_{(\mathbf{D})(\mathbf{N})(\mathbf{C})}) \\
\nonumber &&
+\gamma_{(\mathbf{M})(\mathbf{A})(\mathbf{C})}\gamma_{(\mathbf{B})(\mathbf{N})(\mathbf{D})}-\gamma_{(\mathbf{M})(\mathbf{A})(\mathbf{D})}\gamma_{(\mathbf{B})(\mathbf{N})(\mathbf{C})}],
\end{eqnarray}
where the
$\gamma_{(\mathbf{A})(\mathbf{B})(\mathbf{C})(\mathbf{D})}$ are
the Ricci rotation coeficients.
 We now consider a massive test
particle. Since the particle has a non-vanishing rest mass, it is
convenient to define the tangent vector $V$ as a timelike one, so
that $G(V,V)=-1.$ Let us build a Fermi-Walker transport. In the
Fermi-Walker transported particle frame the equation (2.8) is
given by
\begin{equation}
\frac{d^2Z_{A}}{d\tau ^2}+K_{0A0C}Z_{C}=0,
\end{equation}
where $\tau $ is, in general, an affine parameter, which in our
case is the proper time of the particle and $Z_{{A}}$ are the
vielbein components of the space-like vector $Z$, with $G(Z,V)=0$,
and ${A}$ $\in$ $(1,2,3,4,5,6,\ldots,n)$.

\renewcommand{\theequation}{\thesection.\arabic{equation}}
\section{\bf Modified Hamiltonian Formalism}
\setcounter{equation}{0} $         $

We begin this section with the following first order non-linear
differential equation
\begin{equation}
 \frac{d{y}}{d\tau}=y^{2},
\end{equation}
with the initial conditions in $t=0, y=y_{0},
\frac{d{y}}{d\tau}=y_{0}^{2},$ and solution
\begin{equation}
 y=\frac{y_{0}}{1-y_{0}t}.
\end{equation}
  We present the first order linear differential equation
\begin{equation}
 \frac{d{y}}{d\tau}+p(t)y=g(t),
\end{equation}
and impose that (4.2) is also solution for (4.3). In this case we
have
\begin{equation}
(\frac{y_{0}}{1-y_{0}t})^{2} +\frac{y_{0}}{1-y_{0}t}p(t)=g(t).
\end{equation}

 There is a  great set of functions p(t) and g(t) that
satisfy (4.4). Unfortunately it is necessary to first solve (4.1),
and sometimes a first order non-linear differential equation has
an implicit solution. Considering we have a first or second order
non-linear differential system, is it possible to build a solution
method that transfers the non-linearity  of the non-linear system
to the coefficients of a linear system? The answer is positive
when Hamilton equations can be put in a special form. In this
section we will build this method and in the other ones we will
make some applications. For this we will use the modified
Hamiltonian formalism, which is reduced to the Hamiltonian
formalism when the transformations are canonical or sympletic. It
is well-known that in the Hamiltonian formalism the Hamilton
equations and the Poisson brackets will be conserved only by a
canonical or sympletic transformation. In [4] we have changed the
non-relativistic time-dependent harmonic oscillator [6],[7] to a
general relativistic approach. In the modified-Hamiltonian
formalism only Hamilton equations will be conserved, in the sense
that they will be transformed into other Hamilton equations by a
non-canonical or non-sympletic transformation, and the Poisson
brackets will not be invariant. We now build a modified
Hamiltonian formalism. Consider a time-dependent Hamiltonian
$H({\tau})$ where ${\tau}$ is an affine parameter, in this case,
the proper-time of the particle. Let us define 2n variables that
will be called ${\xi}^j$ with index j running from 1 to 2n so that
we have ${\xi}^j$ $\in$
$({\xi}^1,\ldots,{\xi}^n,{\xi}^{n+1},\ldots,{\xi}^{2n})$ =$(
{q}^1,\ldots,{q}^n,{p}^1,\ldots,{p}^n)$ where ${q}^j$ and ${p}^j$
are coordinates and momenta, respectively. We now define the
Hamiltonian by
\begin{equation}
 H({\tau})=\frac{1}{2}H_{ij}{\xi}^i{\xi}^j,
\end{equation}
where $H_{ij}$ is a  symmetric matrix. We impose that the
Hamiltonian obeys the Hamilton equation
\begin{equation}
\frac{d{\xi}^i}{d\tau }={J}^{ik}\frac{\partial{H}}{\partial{\xi}^k
} .
\end{equation}
The equation (4.6) introduces the sympletic  J, given by
\begin{equation}
\left(%
\begin{array}{cc}
  O & I \\
  -I & O \\
\end{array}%
\right)
\end{equation}
where O and I are the $n \textbf{x}n$ zero and identity matrices,
respectively. We now make a linear transformation from ${\xi}^j$
to ${\eta}^j$ given by
\begin{equation}
  {\eta}^j={{T}^j}_k{\xi}^k,
\end{equation}
where ${{T}^j}_k$ is  a non-sympletic matrix, and the new
Hamiltonian is given by
\begin{equation}
 Q=\frac{1}{2}C_{ij}{\eta}^i{\eta}^j,
\end{equation}
where $C_{ij}$ is a  symmetric matrix. The matrices H, C, and T
obey the following system
\begin{equation}
 \frac{d{{T}^i}_j}{d\tau}+\frac{d{t}}{d\tau}{{T}^i}_k{J}^{kl}X_{lj}=J^{im}Y_{ml}{{T}^j}_k,
\end{equation}
where $2X_{lj}=\frac{\partial{H_{ij}}}{\partial{\xi}^l
}\xi^{i}+2H_{lj}$ and
$2Y_{ml}=\frac{\partial{C_{il}}}{\partial{\eta}^m
}\eta^{i}+2C_{ml},$ t and $\tau$ are the proper-times of the
particle in two different manifolds. We note that (4.10) is a
first order linear differential equation system in ${{T}^i}_k ,$
and it is the response for what we looked for because the
non-linearity in the Hamilton equations were  transferred to their
coefficients. Consider $\frac{d{t}}{d\tau}X_{lj}=Z_{lj}$ and write
(4.10) in the matrix form
\begin{equation}
 \frac{d{T}}{d\tau}+TJZ=JYT,
\end{equation}
where T, Z and Y are  $2n \textbf{x}2n$ matrices as
\begin{equation}
\left(%
\begin{array}{cc}
  T_{1} & T_{2} \\
  T_{3} & T_{4} \\
\end{array}%
\right)
\end{equation}
with similar expressions for Z and Y. Let us  write (4.11) as
follows
\begin{equation}
 \dot{T_1}=Y_{3}T_{1}+Y_{4}T_{3}+T_{2}Z_{1}-T_{1}Z_{3},
\end{equation}
\begin{equation}
 \dot{T_2}=Y_{3}T_{2}+Y_{4}T_{4}+T_{2}Z_{2}-T_{1}Z_{4},
\end{equation}
\begin{equation}
 \dot{T_3}=-Y_{1}T_{1}-Y_{2}T_{3}+T_{4}Z_{1}-T_{3}Z_{3},
\end{equation}
\begin{equation}
 \dot{T_4}=-Y_{1}T_{2}-Y_{2}T_{4}+T_{4}Z_{2}-T_{3}Z_{4}.
\end{equation}
Now consider
\begin{equation}
 \dot{S_1}=Y_{3}S_{1}+Y_{4}S_{3},
\end{equation}
\begin{equation}
 \dot{S_2}=Y_{3}S_{2}+Y_{4}S_{4},
\end{equation}
\begin{equation}
 \dot{S_3}=-Y_{1}S_{1}-Y_{2}S_{3},
\end{equation}
\begin{equation}
 \dot{S_4}=-Y_{1}S_{2}-Y_{2}S_{4},
\end{equation}
and
\begin{equation}
 \dot{R_1}=R_{2}Z_{1}-R_{1}Z_{3},
\end{equation}
\begin{equation}
 \dot{R_2}=R_{2}Z_{2}-R_{1}Z_{4},
\end{equation}
\begin{equation}
 \dot{R_3}=R_{4}Z_{1}-R_{3}Z_{3},
\end{equation}
\begin{equation}
 \dot{R_4}=R_{4}Z_{2}-R_{3}Z_{4}.
\end{equation}
From the theory of  first order differential equation systems [8],
  it is well-known that the system (4.17)-(4.24) has a solution in the
  region where $Z_{lj}$ and $Y_{ml}$ are continuous functions. In
  this case, the solution for (4.10) or (4.11) is given by
\begin{equation}
 {T_1}=(S_{1}a+S_{2}b)R_{1}+(S_{1}d+S_{2}c)R_{3},
\end{equation}
\begin{equation}
{T_2}=(S_{1}a+S_{2}b)R_{2}+(S_{1}d+S_{2}c)R_{4},
\end{equation}
\begin{equation}
{T_3}=(S_{3}a+S_{4}b)R_{1}+(S_{3}d+S_{4}c)R_{3},
\end{equation}
\begin{equation}
{T_4}=(S_{3}a+S_{4}b)R_{2}+(S_{3}d+S_{4}c)R_{4},
\end{equation}
where a,b,c and d are constant $n \textbf{x}n$ matrices, and
substituting (4.25)-(4.28) in (4.8) we will be completed the
mapping among manifolds. In many situations where it is not
possible or easy to put $\frac{d{t}}{d\tau}, X_{lj}, Y_{lj}$ as
explicit functions of one of the two parameters, t or $\tau,$ we
should expand them in series of $\tau$, for example,[8] so that
with the modified Hamiltonian formalism we can map one
differential equation system into another. In this paper we are
interested in mapping among manifolds by the use of Hamiltonians
for Jacobi fields and also for metric tensors. Therefore, if we
associate $H_{ij}$ and $C_{lk}$ with two vielbein curvatures
$K_{0A0C}$ and $R_{0A0C}$ respectively, or $H_{ij}$ and $C_{lk}$
with the metric tensors $G_{ij}$ and ${\Omega}_{ij},$ also from
two different manifolds, we will have built a local map among
manifolds. It is important to note that the same particle has
different proper-times in different manifolds, so that line
elements are not preserved by local non-sympletic maps among
manifolds, in opposition to general relativity where line elements
are preserved by local coordinate transformations. The derivative
$\frac{d{t}}{d\tau}$ increases the difficulty in (4.10), so that
we assume  the condition $\frac{d{t}}{d\tau}=1$. It implies  in a
decrease on mapped regions. The local non-sympletic maps are well
defined for equal proper-times and time intervals. In this paper,
for the same particle in different manifolds with different
proper-times, we only use the proper-time of one of the manifolds,
so that (4.10) assume the following form
\begin{equation}
 \frac{d{{T}^i}_j}{d\tau}+{{T}^i}_k{J}^{kl}X_{lj}=J^{im}Y_{ml}{{T}^j}_k.
\end{equation}
As consequence (4.21)-(2.24) will be simplified. We end this
section calling to mind that in the Hamiltonian formalism the
Poisson bracket will be an invariant only by a canonical or
sympletic transformation so that in this case (4.8) will be
canonical or sympletic  and it can be non-linear. In the modified
Hamiltonian formalism only Hamilton equations will be conserved in
the sense that they will be transformed into other Hamilton
equations by a non-canonical or non-sympletic transformation
(4.8), where (4.10) or (4.29) also will be obeyed, and the Poisson
brackets will not be invariant.
\renewcommand{\theequation}{\thesection.\arabic{equation}}
\section{\bf Modified Formalism and Jacobi Fields}
\setcounter{equation}{0} $         $
 In this section we assume the
condition
\begin{equation}
 \nabla _{V}V=0,
\end{equation}
where $G(V,V)=-1.$ We now identify the hamilton matrices with
Jacobi fields as follow. Let us consider
\begin{equation}
 H({\tau})=\frac{1}{2}H_{ij}{\xi}^i{\xi}^j=\frac{1}{2}(\pi^{t}
 M^{t}\zeta+\zeta^{t}M\pi),
\end{equation}
where ${\xi}^j$ $\in$ $(
{\xi}^1,\ldots,{\xi}^n,{\xi}^{n+1},\ldots,{\xi}^{2n})$ =$(
{\zeta}^1,\ldots,{\zeta}^n,{\pi}^1,\ldots,{\pi}^n)$ and
 $Z_{{A}}={\zeta}^j$ and ${\pi}^j$ are coordinates and momenta,
respectively, and $H_{ij}$ is a real and symmetric matrix given by
\begin{equation}
\left(%
\begin{array}{cc}
  O & M \\
  M^{t} & O \\
\end{array}%
\right)
\end{equation}
where O is the $n \textbf{x}n$ zero matrix, $M^{t}$ is the $n
\textbf{x}n$ transposed matrix of M with $M_{AC}={V}_{C;A}$, and
${A}$,${C}$ $\in$ $(1,\ldots,n)$. Using the Hamiltonian (5.2) in
the Hamilton equation, where $Z_{{A}}$=${\zeta}^j$, we obtain
(3.3). Let us consider
\begin{equation}
 Q=\frac{1}{2}C_{ij}{\eta}^i{\eta}^j,
\end{equation}
where
\begin{equation}
 C_{lk}=\frac{1}{2 },
\end{equation}
for ${l}$,${k}$ $\in$ $(n+1,\ldots,2n)$. We  use  a vielbein basis
related to a local metric G, as usual. The metric tensor
$G_{\Lambda\Pi}$ has signature $(-,+,+,-,-,\ldots,-,+,-),$ with
curved indices $\Lambda,\Pi$ $\in$ $(0,1,2,3,4,5,6,\ldots,n)$ and
it is associated to a local coordinate basis. Let us consider the
connection between the vielbein and the local metric tensor
\begin{equation}
G_{\Lambda\Pi}=E_{\Lambda}^{(\mathbf{A})}E_{\Pi}^{(\mathbf{B})}\eta_{(\mathbf{A})(\mathbf{B})},
\end{equation}
where $ \eta_{(\mathbf{A})(\mathbf{B})}={b_l}{\delta}_{lk},$ and $
E_{\Lambda}^{(\mathbf{A})}$ are a pseudo-euclidian metric and
vielbein components respectively, and flat indices
$(\mathbf{A}),(\mathbf{B}),l,k,\ldots,$ $\in$
$(0,1,2,3,\ldots,n).$ The curvature components in the Fermi-Walker
transported particle frame are given by
\begin{equation}
 C_{lk}=\frac{1}{2}{K_l}{\delta}_{lk},
\end{equation}
 for $ ${l}$,${k}$ \in$ $(1,2,3,4,5,6,\ldots,n)$,
where each ${K_l}$ is a real constant, positive or negative,
${\delta}_{lk}$ is the Kronecker's delta function. Using now the
Hamiltonian (5.4) in the Hamilton equation,  we obtain the
following system
\begin{equation}
\frac{d^2x_{i}}{d\tau ^2}+{K_i}{\delta}_{ij}x_{j}=0,
\end{equation}
where $x_{i}=\eta_{i}$ are the Jacobi fields in the  manifold
given by (5.5) and (5.7). It is a n-dimensional maximally
symmetric space or a pseudo-sphere if ${K_l}=K,$ where K is a
positive constant.
 For a more general case, where each ${K_l}$ is a real constant,
 positive or negative,  we have built a local map between one general
 Riemaniann manifold and another one that is compact in some directions
 and non-compact in other directions. For  ${K_l}$ as a real and positive constant, the
 system (5.8) also appears in the  analysis of the Jacobi
 fields on a n-sphere [9], or geodesic motion on a n-dimensional
 maximally symmetric space-time [10], both  embedded in a (n+1)-dimensional
 Euclidian manifold. In this case the curvature has the following
 expression
\begin{equation}
K_{(\mathbf{A})(\mathbf{B})(\mathbf{C})(\mathbf{D})}=K_{\Lambda
\Pi \Omega
\Theta}E^{\Lambda}_{(\mathbf{A})}E^{\Pi}_{(\mathbf{B})}E^{\Omega}_{(\mathbf{C})}E^{\Theta}_{(\mathbf{D})},
\end{equation}
where
\begin{equation}
K_{\Lambda \Pi \Omega
\Theta}=K(G_{\Lambda\Theta}G_{\Pi\Omega}-G_{\Lambda\Omega}G_{\Pi\Theta}),
\end{equation}
and in the Fermi-Walker transported particle frame (5.9) assume
the simple form
\begin{equation}
 K_{0A0C}=K{\delta}_{AC}.
\end{equation}
We note that (5.11) is a special case of (5.7). Sometimes, as in
[4], is convenient to assume the following Hamiltonian matrix
elements
\begin{equation}
 H_{ij}=K_{0A0C},
\end{equation}
where $K_{0A0C}$ is the Riemannian curvature in a Fermi-Walker
transported particle frame, and  ${i}$,${j}$,${A}$,${B}$ $\in$
$(1,2,3,4,5,6,\ldots,n)$. For ${i}$,${j}$  $\in$ $(n+1,\ldots,2n)$
we have
\begin{equation}
 H_{ij}=\frac{1}{2 }.
\end{equation}
Explicitly,
\begin{equation}
 H({\tau})=\frac{1}{2}H_{ij}{\xi}^i{\xi}^j=\frac{1}{2}
 (P_{A}P^{A}+Z^{A}K_{0A0C}Z^{C}).
\end{equation}
Substituting (5.14) into Hamilton equation  we will obtain (3.3).
Hamiltonians  of type (5.14) are not appropriate for Jacobi fields
on a non-geodesic curve.

\renewcommand{\theequation}{\thesection.\arabic{equation}}
\section{\bf Jacobi Fields on a Non-Geodesic Curve}
\setcounter{equation}{0} $         $
 In this section we assume the condition
\begin{equation}
 \nabla _{V}V\neq0,
\end{equation}
where $G(V,V)=-1.$ In this case the Jacobi equation is not reduced
to a geodesic deviation, and the Fermi derivative
$\frac{D_FZ}{\partial s}$
 does not coincide with the usual covariant derivative
$$
\frac{D_FZ}{\partial s}\neq\frac{DZ}{\partial s}=\nabla _VZ.
$$
We intend to make use of a vielbein basis related to a local
metric G, as in section $3$. The metric tensor $G_{\Lambda\Pi}$
has signature $(-,+,\ldots,+),$ with curved indices $\Lambda,\Pi$
$\in$ $(0,1,2,3,4,5,6,\ldots,n)$ and it is associated to a local
coordinate basis. Let us consider the connection between the
vielbein and the local metric tensor
\begin{equation}
G_{\Lambda\Pi}=E_{\Lambda}^{(\mathbf{A})}E_{\Pi}^{(\mathbf{B})}\eta_{(\mathbf{A})(\mathbf{B})}.
\end{equation}
Considering a  Fermi-Walker transported particle frame and also
the associated Hamiltonian given by  the following function
\begin{equation}
 H({\tau})=\frac{1}{2}H_{ij}{\xi}^i{\xi}^j=\frac{1}{2}(\pi^{t}
 M^{t}\zeta+\zeta^{t}M\pi),
\end{equation}
where ${\xi}^j$ $\in$ $(
{\xi}^1,\ldots,{\xi}^n,{\xi}^{n+1},\ldots,{\xi}^{2n})$ =$(
{\zeta}^1,\ldots,{\zeta}^n,{\pi}^1,\ldots,{\pi}^n)$ and
 $Z_{{A}}={\zeta}^j$ and ${\pi}^j$ are coordinates and momenta,
respectively, and $H_{ij}$ is a real and symmetric matrix given by
\begin{equation}
\left(%
\begin{array}{cc}
  O & M \\
  M^{t} & O \\
\end{array}%
\right)
\end{equation}
where O is the $n \textbf{x}n$ zero matrix, $M^{t}$ is the $n
\textbf{x}n$ transposed matrix of M with $M_{AC}={V}_{C;A}$, and
${A}$,${C}$ $\in$ $(1,\ldots,n)$. Using the Hamilton equation we
have
\begin{equation}
\frac{d{\zeta}^A}{d\tau}=\frac{\partial{H}}{\partial{\pi}^A}={V}_{A;C}\zeta^C,
\end{equation}
and
\begin{equation}
\frac{d{\pi}^C}{d\tau }=-\frac{\partial{H}}{\partial{\zeta}^C }.
\end{equation}
 By the derivative of (6.5) we obtain the following
 result [2]
 \begin{equation}
\frac{d^2\zeta_{A}}{d\tau ^2}+( R_{0A0C}-
\dot{V}_{A;C}-\dot{V}_{A}\dot{V}^{C})\zeta_{C}=0,
\end{equation}
where (6.7) is the Jacobi equation on a non-geodesic curve and
$\dot{V}_{A}={V}_{A;C}V^{C}$. Let us consider a Fermi-Walker
transported particle frame in a new manifold. The associated
Hamiltonian is given by  the following function
\begin{equation}
 Q({\tau})=\frac{1}{2}C_{ij}{\eta}^i{\eta}^j=\frac{1}{2}(\Pi^{t}
 N^{t}\chi+\chi^{t}N\Pi)
\end{equation}
where $g(U,U)=-1,$  ${\eta}^j$ $\in$ $(
{\eta}^1,\ldots,{\eta}^n,{\eta}^{n+1},\ldots,{\eta}^{2n})$ =$(
{\chi}^1,\ldots,{\chi}^n,{\Pi}^1,\ldots,{\Pi}^n)$ and ${\chi}^j$
and ${\Pi}^j$ are coordinates and momenta, respectively, and
$C_{ij}$ is a real and symmetric matrix given by
\begin{equation}
\left(%
\begin{array}{cc}
  O & N \\
  N^{t} & O \\
\end{array}%
\right)
\end{equation}
where O is the $n \textbf{x}n$ zero matrix, $N^{t}$ is the
transpose of the $n \textbf{x}n$  matrix N with
$N_{AC}={U}_{C;A}$, and ${A}$,${C}$ $\in$ $(1,\ldots,n)$. Using
The Hamilton equation we have the following results
\begin{equation}
\frac{d{\chi}^A}{d\tau}=\frac{\partial{Q}}{\partial{\Pi}^A
}={U}_{A;C}\chi^C,
\end{equation}
and
\begin{equation}
\frac{d{\Pi}^C}{d\tau }=-\frac{\partial{Q}}{\partial{\chi}^C }.
\end{equation}
 By the derivative of (6.10) we obtain
 \begin{equation}
\frac{d^2\chi_{A}}{d\tau ^2}+( R_{0A0C}-
\dot{U}_{A;C}-\dot{U}_{A}\dot{U}^{C})\chi_{C}=0,
\end{equation}
where (6.12) is the Jacobi equation on a non-geodesic curve in the
new manifold. Now we can  build a map between the old and the new
manifolds by the non-sympletic linear transformation (4.8). From
the theory of  first order differential equation systems [8], it
is well-known that the system (4.10) or (4.29) has a solution in
the region where the  elements of the matrices $X_{lj}$ and
$Y_{ml}$ are continuous functions. In other words, we have mapped
(6.7) on (6.12).
\renewcommand{\theequation}{\thesection.\arabic{equation}}
\section{\bf Modified Formalism and Metric Tensor}
\setcounter{equation}{0} $         $
 In this section we identify the Hamilton matrices with the
 metric tensors of two different manifolds. Let us  assume that
\begin{equation}
 H_{ij}=G_{ij},
\end{equation}
where $G_{ij}$ is a local metric tensor, and for the new
Hamiltonian we assume the simple case given by
\begin{equation}
 C_{lk}=\frac{1}{2}{b({\tau})_{l}}{\delta}_{lk},
\end{equation}
where ${\tau}$ is an affine parameter, and $C_{lk}$ is a metric
tensor with ${i}$,${j}$,${l}$,${k}$ $\in$
$(1,2,3,4,5,6,\ldots,n)$. For ${i}$,${j}$,${l}$,${k}$ $\in$
$(n+1,\ldots,2n)$ we have
\begin{equation}
 H_{ij}=\frac{1}{2 },
\end{equation}
and
\begin{equation}
 C_{lk}=\frac{1}{2 }.
\end{equation}
Explicitly,
\begin{equation}
 H({\tau})=\frac{1}{2}H_{ij}{\xi}^i{\xi}^j=\frac{1}{2}
 (P_{i}P^{i}+X^{i}G_{ij}X^{j}),
\end{equation}
and
\begin{equation}
 Q({\tau})=\frac{1}{2}C_{ij}{\eta}^i{\eta}^j=\frac{1}{2}
 (p_{i}p^{i}+x^{l}{b({\tau})_{l}}{\delta}_{lk}x^{k}),
\end{equation}
where  $H_{ij}=H_{ji}=G_{ij}=G_{ji},$ and
$C_{lk}=C_{kl}=\frac{1}{2}{b({\tau})_{l}}{\delta}_{lk}$. As each
${b_l}$ can be a positive or  negative function for the same time
interval, we have built a local map between one general
(torsion-free connection or not) manifold and another one that is
compact in some directions and non-compact in other directions. If
$b({\tau})_{l}$ are time-independent $(b({\tau})_{l}=b_{l})$,
then, in this case, the map is equivalent to (3.1). In other
words, it is possible to obtain, with this map, the same results
obtained with vielbein. However, vielbein is also very useful for
Fermi transport which is defined in a torsion-free connection
manifold only. It is possible we chose, as Hamiltonian metric
functions, the usual symmetric forms as follow
\begin{equation}
 H({\tau})=\frac{1}{2}H_{ij}{\xi}^i{\xi}^j=\frac{1}{2}X^{i}G_{ij}X^{j},
\end{equation}
and
\begin{equation}
 Q({\tau})=\frac{1}{2}C_{ij}{\eta}^i{\eta}^j=\frac{1}{2}x^{l}g_{lk}x^{k}.
\end{equation}
We could have substituted (7.5) and (7.6) by (7.7) and (7.8), with
$g_{lk}={b({\tau})_{l}}{\delta}_{lk}$. We note that functions
(7.7) and (7.8) are not  conventional Hamiltonians, but their
associated system (4.17)-(4.24) will be simpler than the
correspondent to (7.5) and (7.6). Now we consider an interesting
Hamiltonian function presented in the last section
\begin{equation}
 H({\tau})=\frac{1}{2}H_{ij}{\xi}^i{\xi}^j=\frac{1}{2}(p^{t}
 M^{t}x+x^{t}Mp),
\end{equation}
where ${\xi}^j$ $\in$
$(({\xi}^1,\ldots,{\xi}^n,{\xi}^{n+1},\ldots,{\xi}^{2n})= (x^{1},
\ldots,x^{n},p^{1},\ldots,x^{n}))$ and
 $ x^{j}$ and $p^{j}$ are coordinates and momenta,
respectively, and $H_{ij}$ is a symmetric matrix given by
\begin{equation}
\left(%
\begin{array}{cc}
  O & M \\
  M^{t} & O \\
\end{array}%
\right)
\end{equation}
where O is the $n \textbf{x}n$ zero matrix, $M^{t}$ is the $n
\textbf{x}n$ transposed matrix of M with $M_{ij}={G}_{ji}$, and
${i}$,${j}$ $\in$ $(1,\ldots,n)$. Now we write another Hamilton
\begin{equation}
 Q({\tau})=\frac{1}{2}C_{ij}{\eta}^i{\eta}^j=\frac{1}{2}(P^{t}
 N^{t}X+X^{t}NP)
\end{equation}
where  ${\eta}^j$ $\in$ $(
{\eta}^1,\ldots,{\eta}^n,{\eta}^{n+1},\ldots,{\eta}^{2n})$ =$(
X^{1},\ldots,X^{n},P^{1},\ldots,P^{n})$ and $X^{j}$ and $P^{j}$
are coordinates and momenta, respectively, and $C_{ij}$ is a real
and symmetric matrix given by
\begin{equation}
\left(%
\begin{array}{cc}
  O & N \\
  N^{t} & O \\
\end{array}%
\right)
\end{equation}
where O is the $n \textbf{x}n$ zero matrix, $N^{t}$ is the
transpose of the $n \textbf{x}n$  matrix N with $N_{lk}={g}_{kl}$,
and ${l}$,${k}$ $\in$ $(1,\ldots,n)$. It is important to note that
in this case $G_{ij}$ and $g_{lk}$ may not have defined
symmetries, and using (4.8) and (4.10) or (4.29) we  transform the
Hamilton equation for (of) $G_{ij}$ into that for (of) $g_{lk}$.
The last map is very important because it connects different areas
both in mathematics and physics.
\section{Concluding Remarks}
  $              $
It is known that the system (5.8) of n-dimensional harmonic
oscillators (for positive ${K_l}$) is the Jacobi equation on a
maximally symmetric spacetime [9], as well as it is the geodesic
equation on a spacetime with constant curvature [10]. For
different sets of signs in ${b_l}$, the system composed by (7.2)
and (7.4) has as solutions different sets of pseudo-euclidian
spaces, where each one of them has form-invariant metric  by a
subgroup of GL(n,R). In Sections 5 and 6 we have presented two
ways of building a map among manifolds by the use of the equations
(4.6), (4.8)and (4.10). Physics of wormholes is a very important
research area [11], [12], [13], [14]. Maps among manifolds can be
thought as wormholes, and can be considered as a different
approach for this. It is also possible to build a traversable
wormhole by a map between two different regions in the same
manifold. Solutions of (4.7) can be very difficult, but in cases
where they are possible, maps among manifolds will be a powerful
option to \emph{conformal maps, Fermi transport, and vielbein
formalism}. Traditionally, expressions like \emph{Hamilton
equations} are associated with some kind of invariance property,
as in the specialized literature. Invariance is a fundamental
property in many theories, as in general relativity. However, if
we want to build maps among manifolds, the modified Hamiltonian
formalism can be useful.


\begin{thebibliography}{99}
\bibitem{1} J. Cheeger and D. G. Ebin, {\bf Comparison Theorems in Riemannian Geometry}
(North-Holland Publishing Company, Amsterdam,1975)
\bibitem{2}  S. W. Hawking and G. F. R. Ellis, {\bf The Large Scale
Structure of Space-Time} (Cambridge University Press,
Cambridge,1973).
  \bibitem{3} C. W. Misner, K. S. Thorne and J. A. Wheeler,
{\bf Gravitation} (W. H. Freeman and Company, San Francisco,
 1973).
\bibitem{4} A. C. V. V. de Siqueira, I. A. Pedrosa, and E. R. Bezerra de Mello {\it
hep-th/9709094v1}.
\bibitem{5} A. C. V. V. de Siqueira, {\it hep-th/0710.1824v1}
\bibitem{6} K. R. Meyer and G. R. Hall, {\bf Introduction to Hamiltonian Dynamical Systems
and the N-Body Problem} (Springer-Verlag, New York, 1991)
\bibitem{7}P. G. L. Leach, {\it J. Math. Phys.} {18},1902,(1977).
\bibitem{8} E. A. Coddington and N. Levinson, {\bf Theory of Ordinary Differential Equations}
(McGraw-Hill, New York, 1955)
\bibitem{9}K. Yano and Y. Muto, {\it Proc. Phys. Math. Soc. Jap.} {18} (1936).
\bibitem{10}S. Weinberg, {\bf Gravitation and Cosmology: princples and applications of the general relativity}(John
Wiley, New York, 1972).
\bibitem{11} M. S. Morris, K. Thorne, and U. Yurtsever, {\it Phys. Rev. Lett.} {61}, 1446, (1988).
\bibitem{12} M. Visser, {\it  Nucl. Phys.} {B328} (1989) 203-212.
\bibitem{13} J. W. Moffat and T. Svoboda,  {\it Phys. Rev.} {D2}, 429, (1991).
\bibitem{14} V. P. Frolov and I. D. Novikov, {\bf Black Hole Physics }
(Kluwer Academic Publishers, Dordrecht, 1998).
\end{thebibliography}
\end{document}